# Design Patterns for Complex Event Processing

Adrian Paschke
BioTec Center, Technical University Dresden,
01307 Dresden, Germany
adrian.paschke AT biotec.tu-dresden.de

**ABSTRACT**
Currently engineering efficient and successful event-driven applications based on the emerging Complex Event Processing (CEP) technology, is a laborious trial and error process. The proposed CEP design pattern approach should support CEP engineers in their design decisions to build robust and efficient CEP solutions with well understood tradeoffs and should enable an interdisciplinary and efficient communication process about successful CEP solutions in different application domains.

## 1. INTRODUCTION

Complex Event Processing (CEP) (see http://www.ep-ts.com/) is an emerging enabling technology to achieve actionable, situational knowledge from distributed systems and data sources in real-time or almost real- time. However, first industrial experiences in using the CEP technology and setting up CEP applications have shown that the early adopters have major problems in adequately engineering successful CEP solutions.

*Design patterns* as more or less formalized descriptions of generic solutions to certain problem classes have become a wide-spread mean to transfer knowledge about successful designs. The advantage of patterns for CEP is their predefined, reusable, and dynamically customizable nature allowing the designer to reuse existing experience for building new CEP solutions. They offer the potential for an additional increase (1) in efficiency, aimed at cheaper and faster delivery of CEP systems for specific domains, and (2) in reusability of successful CEP solutions in various domains.

The multi-dimensional categorization scheme for CEP patterns, proposed in this paper, should enable an efficient communication process about design solutions for various CEP problems and should facilitate building-up comprehensive CEP pattern libraries. Such collected, described and categorized patterns will support CEP application engineers in their design decisions, but will also provide many other use cases as well. For instance, the CEP patterns can be used as a documentation tool, making it easier to understand the tradeoffs of a particular domain-specific CEP solution, open new markets based on exploiting the CEP technology or provide means for an IT team to absorb new CEP developers such as the new role of a *CEP event modeler*, who is responsible for defining the complex event types and situations of relevance.

This paper has to be regarded as a normative proposal. The intention is to stimulate constructive feedback from the pattern community[1], CEP community[2] and closely related communities such as the Reaction Rules community[3] in order to achieve a more general consensus about the proposed CEP architectures and patterns and build-up comprehensive CEP pattern libraries. Clearly, cataloguing, describing and publishing CEP patterns are a community effort.

## 2. Differentiation of Areas

Although CEP is building on the existing fundamental knowledge and the mature methodologies coming from different preceding research fields in computer sciences, necessary groundwork in the very basic definitions of the CEP approach still needs to be done. In this section we contribute with a "by definition" distinction between the more design oriented perspective on CEP addressing CEP models and CEP patterns and the technical view on CEP addressing (complex) event patterns and their processing.

This paper is structured in the following way:

In section 2 a distinction between the design / modelling perspective and the processing perspective is made by a set of fundamental definitions. The main contribution of this paper is the multi-dimensional CEP pattern classification scheme which will be introduced in Section 3. Section 4 present two general pattern language templates for CEP patterns and anti-patterns, introducing the necessary elements which should be commonly included into more specific pattern instantiations of these templates. Finally, in section 5 we conclude this work and give an outline to areas of future work.

### 2.1 Modelling / Design Perspective on CEP

*CEP Model*

A CEP model is a *representation of* a CEP system, whereas systems can be physically observable elements or more abstract concepts like CEP modelling languages.

*CEP Modelling Language*

---

[1] PLoP and EuroPLoP conference series

[2] http://www.ep-ts.com/ and http://complexevents.com

[3] http://ibis.in.tum.de/research/ReactionRuleML

A CEP modelling language is a set of CEP models, and CEP models are *elements of* a CEP modelling language. CEP models *conform to* a CEP model of the CEP modelling language, i.e. a CEP meta-model.

*CEP Megamodels*

CEP models about modelling are called CEP megamodels. A CEP megamodel describes a CEP model as a system that enables us to give answers about a CEP system under study without the need to consider this CEP system directly.

*CEP Metamodel*

CEP meta-models are models of CEP modelling languages. They can be used to validate CEP models. For one modelling language multiple CEP (meta)models can exist, which can differ in the language they are described in.

*Model Transformations*

Model transformations are specified between metamodels. The execution of a model transformation transforms models conforming to the source meta-model into models conforming to the target metamodel. *Vertical model transformations* refine abstract models to more concrete models, while *horizontal model transformations* describe *mappings* between models of the same abstraction level.

*CEP Reference Model*

A CEP reference model is an abstract representation of the entities and relationships involved in a problem space. It forms the conceptual basis for the development of more concrete CEP models of the space, and ultimately CEP implementations, in a concrete application/computing context by customizing the CEP reference model to a particular usage context.

*Best Practices*

Best Practices are a description of successful techniques, methods, processes, activities, incentives or rewards that are more effective at delivering a particular outcome than any other techniques, methods, processes, etc. for a particular domain of interest.

*CEP Patterns* and *CEP Pattern Languages*

CEP Patterns capture and formally codify good designs and best experience-based best practices in a CEP pattern language based on a common vocabulary in such a way that it is possible for others to reuse them. They successfully convey insight into common problems and their solutions. Different categories of CEP patterns can be distinguished (see section 5 for more details). There is also a distinction between a CEP pattern, as a didactic mean to enable actors to behave like an expert, and an event pattern definition, as a description of the detection conditions of a (complex) event (see event pattern).

## 2.2 Processing Perspective on CEP

*Event Pattern*

An event pattern definition (event definition or event type) describes the structure of an (atomic or complex) event, i.e. it describes its internal structure and detection condition(s).

*Event Instance*

A concrete *instantiation* of an event pattern is a specific event instance (also event object).

*Complex Event Processing and Event Processing Languages*

Complex event processing describes the process of event selection, aggregation, hierarching, event abstracting and composing of complex events from raw events for generating higher level events of interest. Event Processing Languages support the specification of event patterns / event definitions, selection and consumption policies, as well as the rules for event processing.

## 3. Categorization Scheme for CEP Patterns

The categorization scheme proposed in this section forms the basis for clustering CEP patterns into vertical domain-specific and generic horizontal across-the-domain dimensions.

### 3.1 Categorization according to Good and Bad Solutions

The first category distinguishes between successful CEP patterns and "bad" CEP anti-patterns describing inefficient solutions.

*CEP patterns*

CEP patterns document a successful solution to a frequently occurring problem.

*CEP Anti-pattern*

CEP anti-patterns are conceptually similar to CEP patterns in that they document recurring solutions to common design problems. They are known as *anti-patterns* because their use (or misuse) produces negative consequences. Anti-patterns document common mistake made during CEP development as well as their solutions.

Note: In the following when we speak of patterns we always mean both patterns and anti-patterns.

### 3.2 Categorization according to the Abstraction Level

The second category distinguishes between the levels of abstraction reaching from CEP architectural design, to concrete development, deployment and optimization patterns, as well as to CEP application management.

*Guidelines and Best Practices*

More or less informally described guidelines and best practices for the design, development, deployment, and management of CEP applications.

*Management patterns*

Management patterns address the management of CEP applications, i.e. they adopt general IT Service Management (ITSM) and business process management (BPM) solutions and best practices to the domain of CEP applications.

*Architecture patterns*

CEP Architecture patterns are high level patterns which describe the general architecture of CEP systems and the interplay of their components and provided features.

*Design patterns:* Design Patterns codify successful proven practice for refining the components and subsystems of a CEP application within a specific context, by more or less formalized documentation of the assumptions, structure, dynamics, applicability and consequences of design decisions, as well as possible design alternatives.

*Mapping patterns*

Mapping patterns combine and efficiently tailor successful design patterns to a concrete CEP product / application. Typically, these (product) mappings are based on proven implementations.

*Idioms / Realization patterns*

Idioms are common practice realization patterns on the technical implementation level. An idiom guides the assembly and implementation of CEP components; often based on the specifics and feature of a concrete event processing language (EPL) or CEP engine.

*Smells*

Smells describe symptoms that indicate that something may be wrong in the specific structures or sub-parts in a concrete technical CEP implementation and should be refactored or the overall design should be reexamined. For instance, wrong, incomplete, or inefficient structures in the definition of complex events that can be improved by the application of *refactoring*. The definition of smells is generally relatively informal as compared to [anti-] patterns.

*Refactoring Patterns*

Refactorings are transformations to improve the quality of a CEP-based solution/implementation, in particular on the technical (code) layer, i.e. on the level of smells and idioms, e.g. the concrete optimization of a complex event pattern.

### 3.3 Categorization according to the Intended Goal

The third categorization distinguishes CEP patterns according to their intended goal, mainly from the view of a CEP solution provider, i.e. which kind of problems in employing the CEP technology should be solved by the pattern.

*Adoption patterns*

Adoption patterns document strategic decisions which speed-up or ease (respectively delay or hinder) the adoption of CEP solutions and tools by business and customers.

*Business patterns*

Business patterns describe successful end-to-end CEP business applications and identify the involved businesses partners, customers, and their interactions.

*Integration patterns*

Integration patterns describe feasible combinations of business patterns in order to create CEP applications with added value and advanced functionality.

*Composite patterns*

Composite patterns are combinations of business patterns and integration patterns that have themselves become commonly used types of CEP applications. Composite patterns are advanced CEP applications.

*Workflow patterns*

Workflow/Process patterns define the concrete process flow in a CEP system or application, hence are concrete specifications of business processes (business patterns) and/or application workflows (integration and composite patterns).

*Coordination patterns*

Coordination patterns partially overlap with workflow and process patterns. But where such process or workflow patterns describe the control flow of the business or CEP application logic, the coordination patterns focus on the different points of the interaction between components in a CEP business process, i.e. describe successful coordination protocols.

*Customized patterns*

Customized patterns relate to composite patterns, as they combine integration patterns and business patterns to form an added value, end-to-end solution. However, they only provide solutions to solve problems of one specific company within a specific context.

*Application patterns*

Application patterns describe the implementation of concrete CEP applications that fulfill certain customer's requirements. They specify the existing CEP technologies and supporting runtime environments.

### 3.4 Categorization according to the Management Level

Finally, the last category makes a general classification of CEP patterns into strategic patterns, tactical patterns and operational pattern, i.e. they describe design or management decisions on the operational, tactical and strategic level of CEP application/service management.

*Strategic patterns*

Strategic pattern, or CEP business value management patterns, describe the strategic alignment of the CEP-based IT into the long-term business strategy. They are an integral part of the enterprise governance and describe successful leadership and organizational structures and processes that ensure that the organization's CEP infrastructure sustains and extends the organization's strategy and objectives. They are part of the general IT governance strategy of an enterprise.

*Tactical patterns*

This type of patterns superimposes the management patterns and describes best practices for (business) processes that cooperate to provide added value and ensure persistent quality of the CEP-based applications to the customer. Typically such processes are based on existing solutions in IT Service Management (ITSM, such as service level management, change management, asset management and problem management), business activity monitoring (BAM), and business process management (BPM).

*Operational patterns*

Operational patterns focus on optimizing the management of the CEP application infrastructure, i.e., the components it contains and the data it creates. They build on IT infrastructure management (ITIM) and the operational processes in ITSM.

### 3.5 Multi-dimensional mapping of CEP pattern categorizations levels

Based on these three categories[4] we can derive a multi-dimensional categorization scheme as shown in figure 1. CEP patterns can be categorized into this scheme which reveals connections and dependencies between the three dimensions of CEP pattern types.

## 4. CEP Pattern Language

Many different design pattern languages have been introduced in the past two decades in various disciplines (see e.g. proceedings of the major PLoP conferences or (Gamma 1995, Hillside.net)). It is not within the scope of this paper to give a comprehensive overview on them and discuss their merits. The two presented general templates for CEP pattern languages, adapted from (Gamma 1995, Hillside.net), shall introduce a common core of shared descriptive elements. These elements should be included into more specific instantiations of these templates, independent of what particular notational representation formalisms are used.

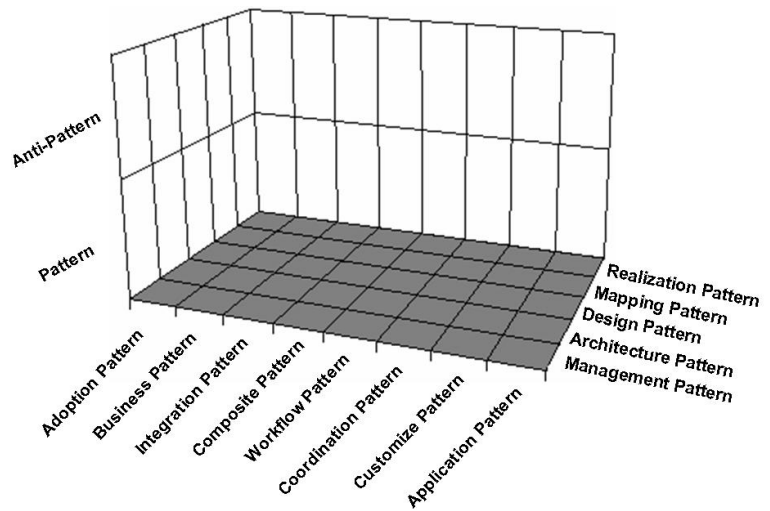

**Figure 1: Three-dimensional categorization scheme for CEP patterns**

**Pattern language template:**

*Name*
A name used for identification

*Problem*
A repeating problem that occurs in a domain

*Solution*
Best practice solution to that problem

*Consequences*
Advantages and disadvantages of the recommended solution

*Examples*
A few examples where the recommended solution has already been applied

**Anti-pattern language template:**

*Name*
A succinct name to convey the essence of the anti-pattern

---
[4] We omit the operational, tactical and strategic categorization level here since it superimposes the other categories

*Problem / Bad solution*

The commonly occurring mistake or bad solution that relates to the anti-pattern

*Symptoms*

The indications or signs of the problem

*Consequences*

The results of applying this anti-pattern

*Root cause*

This provides the context for the anti-pattern, that is, where a pattern was applied incorrectly and resulting in a problem or failed solution

*Suggested solution(s)*

Refactored solution that solves the problem and ensures more benefits

## 5. Conclusion and Future Steps

In this paper we have presented a classification scheme to further evolve a pattern based engineering approach for CEP applications in a more structured way. This is a first step to make the CEP technology tractable by easy-to-use methods, technologies and tools, and to provide integrated solutions and best practices to practitioners in major industry sectors.

As part of a larger integrated project on "Domain-specific Reference Models for CEP Patterns" (DoReMoPat), we currently build up a comprehensive online library of CEP patterns[5] which should enable practitioners and researchers to communicate effectively about successful domain-specific CEP solutions. This involves three important areas of current research:

1. Develop an adequate CEP pattern language and use it to determine, describe and categorize best practices and successful CEP solutions according to the pattern categorization scheme introduced in this paper. This should lead to a detailed and comprehensive library of domain-specific and across-the-domain CEP reference architectures, reference models and patterns.
2. Define typical design criteria and implement a rule-based decision support system on top of the online CEP pattern library that supports engineers in their design decisions, i.e. a service that allows a designer choosing the "right" pattern for a given business and CEP application context.
3. Significant efforts are necessary to come up with a (semi-)formal specification/modelling framework facilitating the (semi-) automated generation of new CEP applications by customization of reference architectures and models, and their solution-oriented design pattern specifications into the context of an application domain;

Our goal is to initiate a community effort and stimulate a constructive feedback from the CEP community[6] in order to achieve a more general consensus about CEP patterns and the used terms and concepts. In the end, we will contribute with a comprehensive collection of CEP patterns together with a helpful toolbox aimed at improving the efficiency of the CEP application engineering process and the quality of CEP-based solutions.

**References**

Hillside.net, *Design Patterns Homepage*
*http://hillside.net/patterns/*

Gamma, E., et al., *Design Patterns - Elements of Reusable Software*. 1995: Addison-Wesley

---

[5] see e.g., Ammon, R. v., Silberbauer, C., Wolff, C. Domain Specific Reference Models for Event Patterns – for Faster Developing of Business Activity Monitoring Applications. VIPSI 2007 Lake Bled, Slovenia, 8-11 October 2007

[6] http://www.ep-ts.com/ and http://complexevents.com/